\documentclass[11pt,a4paper]{article}
\usepackage{graphicx}% Include figure files
\usepackage{dcolumn}% Align table columns on decimal point
\newcommand{\be}{\begin{equation}}
\newcommand{\ee}{\end{equation}}
\newcommand{\bea}{\begin{eqnarray}}
\newcommand{\eea}{\end{eqnarray}}

\begin{document}

\setcounter{page}{1}

\title{Nuclear modification and azimuthal anisotropy
of D mesons produced in relativistic heavy ion collision}
\author{Mohammed Younus}
\date{}
\maketitle
\begin{center}
\it{Bose Institute, 80/EN, Bidhan Nagar, Kolkata,
 700 091, India}
\end{center}

\begin{abstract}
In this paper we present a phenomenological
treatment of charm quark energy loss
before fragmenting into D mesons
and calculate nuclear modification
factor, '$R_{AA}$' and azimuthal anisotropy, '$v_2$'
of D mesons for lead on lead collision at LHC
 energy of $\sqrt{s}$=2.76 A TeV.
\end{abstract}

%\pacs{11.10.Wx}

%\keywords{}
\maketitle

\section{Introduction}
Whenever two heavy ions collide with ultra-relativistic
energies and pass through each other, there
remains a region at their
point of collision where we have a system of partons,
composed of gluons, and some quarks and anti-quarks in
apparently deconfined state but within a very small volume
in space which is larger than the average
nucleon sizes. These quarks and gluons
soon relegate to their lower energy states via elastic
and inelastic collisions and reach an equilibrium
momentum distribution. This locally thermal
equilibrated system is known as quark-gluon-plasma (qgp)
~\cite{qgp1}--\cite{qgp3}. Investigating the properties of
quark gluon plasma remains a major activity
of present day high energy nuclear physics and
promises a deeper understanding of the laws
of quantum chromo-dynamics (QCD). Suppression
of hadrons~\cite{sig1}, heavy quarks~\cite{sig2},
jet-quenching~\cite{sig3}, radiation of thermal photons~\cite{sig4},
suppression and regeneration of $J/\psi$~\cite{sig5} are some
of the observables which we can evaluate as the
signatures for the formation of qgp. To start with our discussion
let us look briefly into the heavy quark production
in relativistic heavy ion collision.

The production rate for the massless gluons and
lighter quarks can be traced throughout heavy ion
collision phases as a minimal momentum transfer is
required for their production.
On the other hand owing to large mass, charm quarks are believed to be
mostly produced in
the pre-equilibrium phase of relativistic heavy ion collision
, where
partonic momenta are relatively very high. The
rate of production of charm quark is limited
in the later phases of collision history~\cite{pro1},
as temperature of the thermal medium is far below
the charm quark mass.
Thus being separated from the bulk of qgp,
and due to its small numbers, heavy quarks
can serve as probes to qgp properties. As expected, heavy
quarks like other probe particles 
undergo collisional and radiative energy loss due
to scattering with the medium partons and this is observed
as a suppression in final heavy meson spectrum
via nuclear modification factor, '$R_{AA}$'~\cite{suppress1,suppress2}.
Evidently recent data from RHIC and
LHC~\cite{dat1,dat2} experiments show large suppression for heavy
mesons which is almost identical to lighter mesons
and against the general belief that
heavy quark may not loose energy comparable
to gluons and light quarks.
Thus studying heavy quark energy loss
mechanisms in thermal bath and
calculation of nuclear modification
factor, '$R_{AA}$', have emerged as
contemporary topics and will also
be featured in the present literature.

Next let us look into another observable,
heavy quark azimuthal anisotropy, $v_2$ also
featured in this paper. Data
observed for non-central collisions both at RHIC and LHC, exhibits
considerable elliptic flow particularly for the low momentum
D mesons~\cite{dat3}, and is found to be comparable to
lighter mesons, $v_2$. However
one must be careful not to mix heavy mesons results with
medium partons' elliptic flow unless
heavy quark is assumed thermalized
and cannot be distinguished from bulk medium any longer.
This is however contrary to the conclusions that heavy quark
may not fully thermalize in quark gluon plasma unless it
interacts with the medium strongly~\cite{therm1}. On the other hand
with the medium temperatures reaching 400-500 MeV at
LHC collider energies, it is predicted that low momentum
charms may thermalize within the given life time of qgp,
which may explain D mesons' large elliptic flow at
low transverse momentum sector.

Thus along with nuclear suppression and elliptic
flow, the study of heavy quark evolution in
quark gluon plasma presents an unique and contemporary field of
research.
There have been some recent theoretical and phenomenological
studies of heavy quark dynamics in the thermal medium
most notably~(\cite{hqw1}--\cite{hqw4}), where along with $R_{AA}$, and $v_2$,
other observables such as
heavy quark correlation, $C(\Delta\phi, p_T, y, \eta)$
in transverse momentum, azimuthal angle and rapiditites
~\cite{hqc1},
heavy quark energy loss per unit path-length in medium, $dE/dx$,
and transport coefficients, such as momentum broadening,
$\hat{q}$, drag coefficient, $A_i(p)$ and diffusion
coefficient, $B_{ij}(p)$, heavy quark thermalization
rates have been calculated in detail.
The temperature, momentum and the path length dependence
of the above transport coefficients using
transport models, at both leading order and
next-to leading order feynman diagrams
using resummed loop diagrams calculations in
hard thermal loop approximation has been calculated, to get a
more rigorous picture of heavy quark evolution dynamics
in quark gluon plasma.

In this paper we present a phenomenological treatment
of charm quark energy loss and calculate nuclear modification
factor, '$R_{AA}$' and azimuthal anisotropy, '$v_{2}$' of D
mesons, at LHC $\sqrt{s}$= 2.76 A TeV. The recent
data from ALICE has shown D mesons, '$v_2$'
and '$R_{AA}$' for various centrality
classes. This paper tends to
explain the data using collisional and
radiative energy loss models. Over the
few following sections, we would discuss about charm
production, charm energy loss, calculations of
the above mentioned
observables, followed by discussions of our results and
conclusions.

\begin{figure}
%\begin{center}
\includegraphics[scale=0.30,angle=270]{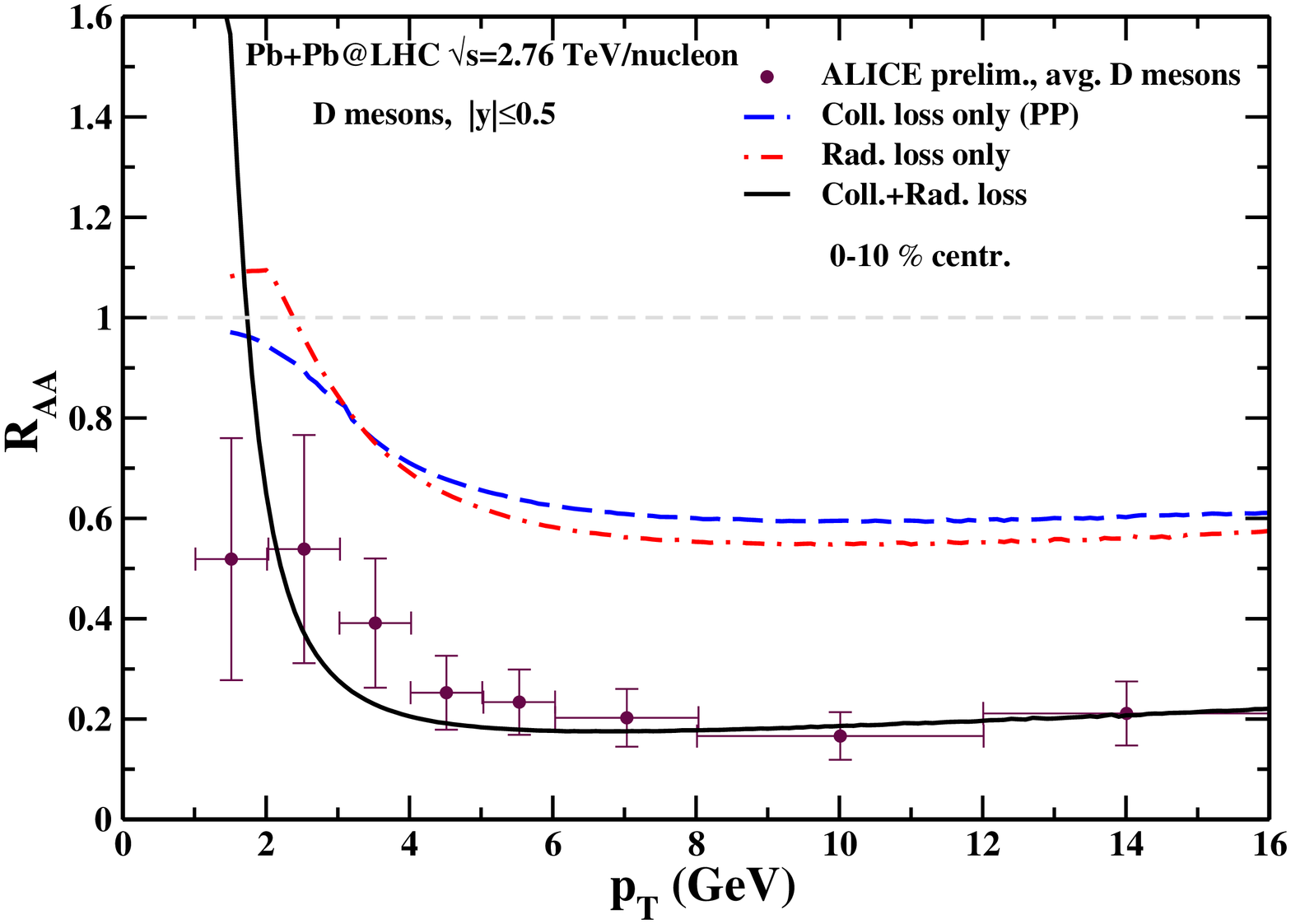}
\includegraphics[scale=0.30,angle=270]{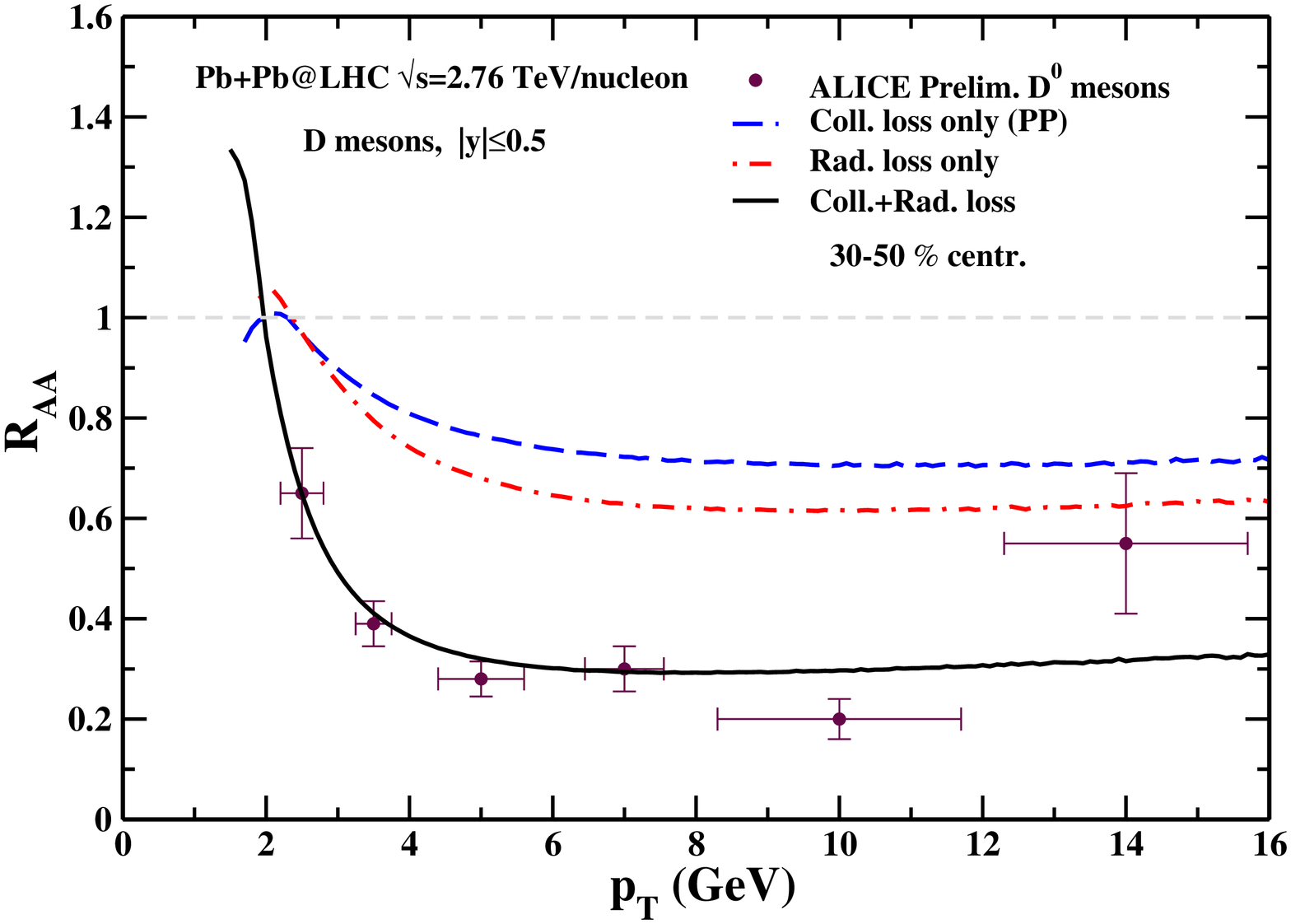}
\includegraphics[scale=0.30,angle=270]{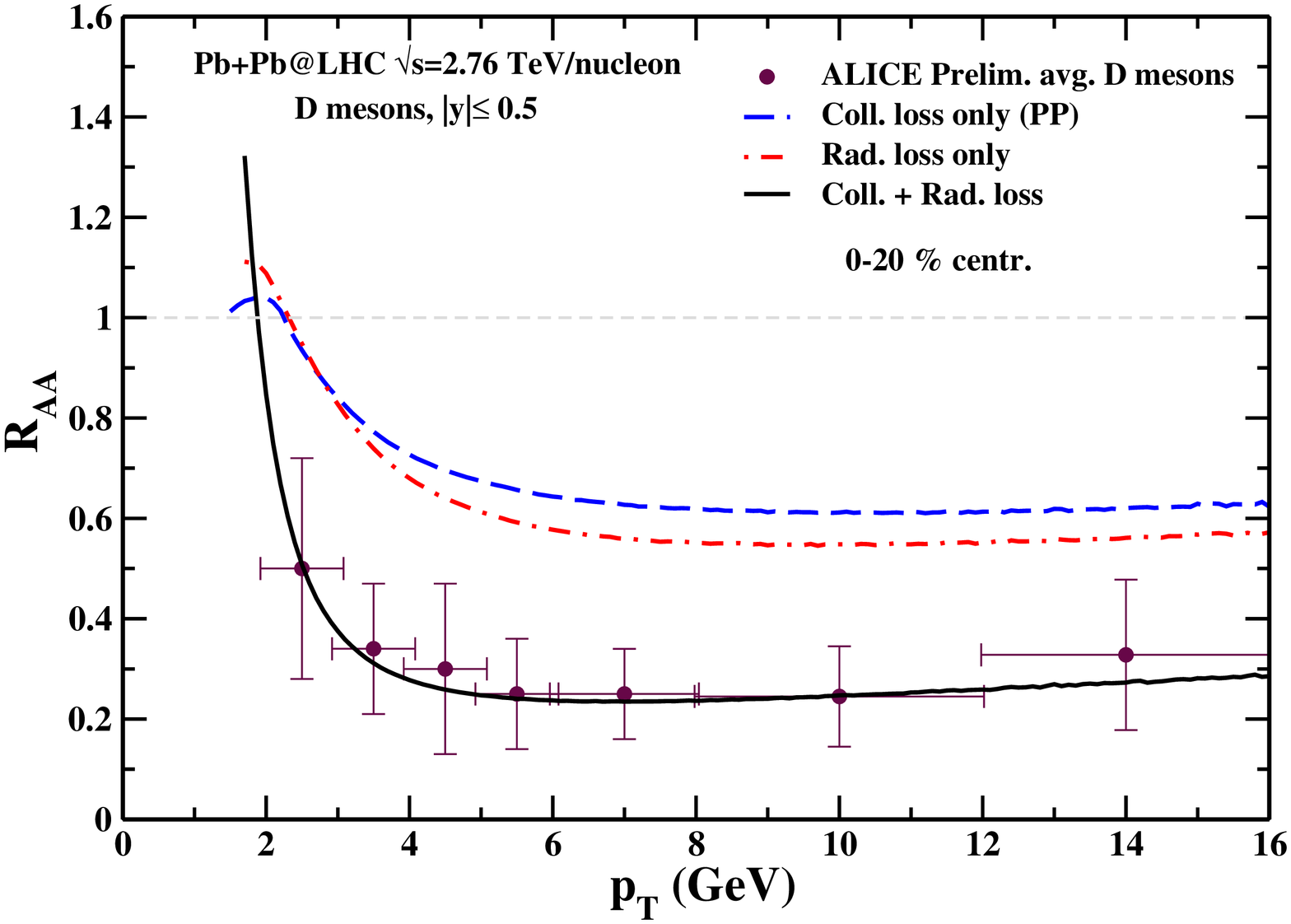}
%\end{center}
\caption{Nuclear modification factor, '$R_{AA}$',(top left)0-10\% centr.,
(top right)0-20\% centr. and (bottom)30-50\% centr.}
\label{fig1raa}
\end{figure} 

\section{Charm Production}

As discussed in the previous section, charm quark
is produced mostly in the pre-equilibrium phase
of heavy ion collision and its production
cross-section can be approximately
scaled from proton on proton collision using
nuclear overlap function, '$T_{AA}(b)$', where '$b$',
commonly called impact parameter,
is the separation of the centres of two
colliding nuclei and denotes centrality class
of the collision.
One can now calculate charm production at
%J.~D.~Bjorken,
  %``Highly Relativistic Nucleus-Nucleus Collisions: The Central Rapidity Region,''
%  Phys.\ Rev.\ D {\bf 27}, 140 (1983).
the leading order(LO), in '$AA$' collision as~\cite{pro1,pro2},
\begin{equation}
\frac{dN^{c}_{AA}}{d^2p_{T}dy}=T_{AA}(b)\frac{d\sigma^{c}_{pp}}{d^2p_Tdy}
\label{eq1}
\end{equation}
in which
\begin{eqnarray}
\frac{d\sigma_{pp}^c}{d^{2}p_{T}dy} &=& 2 x_{a}x_{b}\sum_{ij}
\left[f^{(a)}_{i}(x_{a},Q^{2})f_{j}^{(b)}(x_{b},Q^{2})
\frac{d\hat{\sigma}_{ij}(\hat{s},\hat{t},\hat{u})}{d\hat{t}}
\right.\nonumber\\
&+& \left.f_{j}^{(a)}(x_{a},Q^{2})f_{i}^{(b)}(x_{b},Q^{2})
\frac{d\hat{\sigma}_{ij}(\hat{s},\hat{u},\hat{t})}{d\hat{t}}\right]
/(1+\delta_{ij})~,
\label{eq2}
\end{eqnarray}
where $p_{T}$ and $y_{1,2}$ are the momenta and rapidities of
produced charm and anti-charm and
$x_{a} $ and $x_{b} $ are the fractions of the momenta carried by the
interacting partons
from their respective parent hadron/nucleon. $f_{i/j}(x,Q^2)$
are the partonic distribution functions inside the nucleons.
We have used CTEQ5M partonic structure function~\cite{pdf1} with EKS98
shadowing function~\cite{sdf1} for our parton distribution function,
and a constant strong coupling, $\alpha_s\simeq$0.3 for our calculations.

We have used  $T_{AA}(b)$= 277 fm$^{-2}$ for 0-10$\%$ centrality,
260 fm$^{-2}$ for 0-20$\%$ centrality, and
40 fm$^{-2}$ for 30-50$\%$ centrality
for Pb+Pb at LHC, $\sqrt{s}=$
2.76 TeV/nucleon calculated from Glauber formalism.

The $p_{T}$ distribution of charm production is calculated
using Eq.~\ref{eq2} at the leading order(LO) and a $K$-factor of 2.5 is taken to include
the next-to-leading order effects. The fundamental processes included
for LO calculations of invariant scattering amplitudes~\cite{pro3},
$|M|^2$ are $g+g \rightarrow c+\overline{c}\,;q+\bar{q} \rightarrow c+\overline{c}$
and finally the differential scattering cross-section
to be used in Eq.~\ref{eq1} is given by,
\begin{eqnarray}
\frac{d\hat{\sigma}}{d\hat{t}}=\frac{1}{16\pi(\hat{s}-m_c^2)^2}|M|^2
\label{eq3}
\end{eqnarray}

We can now move over to calculation
of the energy loss mechanism of the produced charm quarks.

\section{\bf {Medium effects on charm}}

\subsection{Collisional energy loss}
The charm quark on entering quark gluon plasma
undergoes collision with the medium partons with
the rate, $\Gamma\sim\frac{1}{\lambda}\sim\sigma_{2\rightarrow 2}
\rho_{qgp}$. '$\lambda$' is the mean free path between two
successive collision, '$\sigma_{2\rightarrow 2}$' is the
total elastic scattering cross-section and '$\rho_{qgp}$'
is the qgp medium density, respectively. The scattering
processes in our calculations involve
a small momentum transfer, '$q_\perp^2$'
that slows down the charm quark. Thus eikonal approximation
for the probe charm quark energy, $E\gg q_\perp$ is maintained throughout
and $2\rightarrow 2$ scattering processes are calculated
within the framework of pQCD assuming
a constant value of 0.3 for strong coupling, '$\alpha_s$'.

The assumptions taken above may not be true for all charm
energies particularly for low energy sector where non-perturbative
effects may play a vital role. Similarly higher order corrections
and loop diagrams resummation techniques also play important roles as
demonstrated earlier by~\cite{coll1} . We will come back to this later.

One can approximtely show collisional energy
loss per unit length traveled by charm quark in the medium as,

\begin{eqnarray}
\frac{dE_{coll}}{dx}=\sum_{i}{\frac{1}{8vE}
\int{d^3kd^3k'}\frac{n_in_i'}{kk'E'}
\delta^4(p^\mu)|M|^2_{2\rightarrow 2}\varepsilon}
\label{eq4}
\end{eqnarray}
where $n_i$ are thermal medium parton distribution with
3-momentum, '$k$', $E$ and $E'$ are the energies of charm before and
after scattering, and 
'$\varepsilon=E-E'$' is the energy loss in a collision.

The above expression can be simplified using techniques(see ref.~\cite{coll2})
assuming that $E,E'\gg T$, where $T$ is
the temperature of the medium. The final
integration looks like;

\begin{equation}
 \frac{dE_{coll}}{dx}=\sum_i d_i \int\frac{d^3k}{2k}
 \int^{tmax}_{tmin}d\hat{t} (-\hat{t})
 \frac{d\hat{\sigma}}{d\hat{t}}
 \label{eq5}
\end{equation}
One can now calculate analytically to get the expression by A. Peshier et al., 
~\cite{coll2} to obtain $dE_{coll}/dx$.

The leading-order(LO) scattering processes considered here
for differential scattering cross-section, $d\hat{\sigma}/d\hat{t}$,
are $gc\rightarrow gc$ and $q(\bar{q})c\rightarrow q(\bar{q})c$ and
can be calculated or can be taken from~\cite{pro3}.

\subsection{Radiative energy loss}
As discussed in the previous section, the charm
quark upon entering the hot and dense medium
undergoes a series of scattering before
coming out of the medium as D meson. Upon
each of the scattering the charm may acquire
some virtuality which it looses via radiation
of gluons. Radiative loss has
emerged as one of the prominent energy loss
mechanisms of probe charm quark. Radiative
energy loss along with collisional loss is
able to explain experimental data satisfactorily.
However one must be careful while comparing with 
various models available for heavy quark radiation
mechanisms as each of the models
has its own set of cuts and constraints imposed in their
respective calculations (see~\cite{rad1,rad2}).

Now let us return to the present calculation.
Each radiated gluon has its formation time~\cite{form1} after
which it is separated from the radiating
probe particle and is given by
\begin{equation}
 \tau_f=\frac{2\omega}{k_\perp^2}
 \label{eq6}
\end{equation}
If '$\tau_f$' becomes larger than the , '$\lambda$',
which is the mean free path or
the average distance between two successive collisions, then
the radiations from successive scattering interfere
destructively commonly called Landau-Pomeranchuk-Migdal
effect~\cite{LPM1}. As the result the gluon spectrum
from single scattering gets suppressed.
Following the earlier developments by (see Ref.~\cite{DK1}) along with
the hierarchy $E,E^/\gg q_\perp >\omega\gg k_\perp > \mu_D$,
I have heuristically assumed the
gluon spectrum from single scattering to be suppressed by
no. of coherent centres, '$N_{coh}=\frac{\tau_f}{\lambda}=\sqrt{\frac{\omega}{\hat{q}L^2}}$',

\begin{eqnarray}
\frac{dn_g}{d\omega d\eta}=\frac{dn_g^{incoh}}{d\omega d\eta}.\frac{1}{N_{coh}}\,,\nonumber\\
\frac{dn_g}{d\omega d\eta}=\frac{C_A\alpha_s}{\pi}.
\frac{1}{\omega^{3/2}}.\sqrt{\hat{q}L^2}.D(\eta)
\label{eq7}
\end{eqnarray}
where
\begin{equation}
\frac{dn_g^{incoh}}{d\omega d\eta}=\frac{C_A\alpha_s}{\pi}\frac{1}{\omega}D(\eta),
\end{equation}
is the simple Gunion-Berstch formula~\cite{gb1} for gluon radiation from single scattering
centre,
and '$D(\eta)$' is the dead-cone factor taken from~\cite{abir1},
$\omega$ and $\eta$ are the energy and rapidity of the
emitted gluon,
'$L$' is the average path length traveled by charm and
'$\hat{q}$' is the average momentum broadening which can be
calculated as
\begin{equation}
 \hat{q}=\rho\int{q^2_\perp \frac{d\sigma}{dq^2_\perp}}dq^2_\perp
 \label{eq8}
\end{equation}
and is roughly found to be of the '$\mathcal{O}$(1GeV$^2$/fm)'~\cite{hqw3,DK1,hqw4,hqw5}.
$C_A$ is the colour factor taken to be $N_c=$3.0.
The coupling strength, '$\alpha_s$' is taken a constant value of
0.3 throughout our calculations.

Now the radiative energy loss per unit length traveled
by the charm can be written naively,
\begin{equation}
 \frac{dE_{rad}}{dx}=\frac{\langle\omega\rangle_g}{\langle\lambda\rangle}\,,
 \label{eq9}
\end{equation}

Where the average radiated energy by charm is,
\begin{equation}
 \langle\omega\rangle_g=\frac{\int{d\omega d\eta ~
 \omega ~ dn_g/d\omega d\eta}}{\int{d\omega d\eta ~
 dn_g/d\omega d\eta}}
 \label{eq10}
\end{equation}

And the average mean free path can be written as
\begin{equation}
 \langle\lambda\rangle^{-1}=\rho_{qgp}\sigma_{2\rightarrow 3}
 \label{eq11}
\end{equation}
where
\begin{equation}
\sigma_{2\rightarrow 3}=\int_{t_{min}}^{t_{max}}{d\hat{t}\frac{d\hat{\sigma}_{2\rightarrow 2}
}{d\hat{t}}}
\int_{\omega\eta}{\frac{dn_g}{d\omega d\eta}D(\eta)}
\end{equation}

Putting the above equations
together and after a brief simplification
one may finally write eqn. (9) as ,

\begin{equation}
 \frac{dE_{rad}}{dx}=\frac{C_A}{\pi}\alpha_s^3\rho_{qgp}\sqrt{\hat{q}L^2}.
 \int_{t_{min}}^{t_{max}}{d\hat{t}\frac{d\sigma}{d\hat{t}}}
 \int_{\omega_{min}}^{\omega_{max}}{d\omega\frac{1}{\omega^{1/2}}}
 \int_{\eta_{min}}^{\eta_{max}}{d\eta D(\eta)}
 \label{eq12}
\end{equation}

The limits of the integration
for the scattering processes above
are calculated to show;
\begin{eqnarray}
t_{min}=\omega_{min}^2=\mu_D^2=4\pi\alpha_s.T^2\nonumber\\
t_{max}=\frac{3ET}{2}-\frac{M^2}{2}
+\frac{M^4}{64p_TT}\ln{\left[\frac{M^2+6ET+6p_TT}{M^2+6ET-6p_TT}\right]}\nonumber\\
\omega_{max}=\left[\int_{t_{min}}^{t_{max}}{d\hat{t}.\hat{t}.\frac{d\hat{\sigma}}{d\hat{t}}}\right]^{1/2}\nonumber\\
\eta_{min}/\eta_{max}=(-/+)\ln{\left(\frac{\omega_{max}}{\mu_D}
+\sqrt{\frac{\omega_{max}^2}{\mu_D^2}-1}\right)}
\label{eq13}
\end{eqnarray}

Next we can assume as first approximation that
the 'total energy loss per unit path-length'
is simple addition of collisional and radiative terms
shown above
and is given by,
\begin{equation}
 \frac{dE_{tot}}{dx}=\frac{dE_{coll}}{dx}+\frac{dE_{rad}}{dx}
 \label{eq14}
\end{equation}

The average total energy loss, '$\Delta E$' and
total momentum loss, '$\Delta p_T$', are then
calculated over average path length, $\langle L \rangle$,
 traveled by charm in the transverse direction~\cite{pro2}
 and can be shown as;

\begin{eqnarray}
 \langle L \rangle=\frac{\int{drr}\int{d\phi L(r,\phi)T_{AA}(r,b)}}
 {\int{drr}\int{d\phi T_{AA}(r,b)}}\nonumber\\
 L(r,\phi)=-r\cos{\phi}+\sqrt{R^2_b-r^2\sin^2{\phi}}
 \label{eq15}
\end{eqnarray}
where $r\,,\phi$ are the position and angle of motion for the charm
inside the medium on the transverse plane.

We have assumed only temperature variation of the qgp like medium
assuming Bjorken longitudinal expansion only~\cite{bjorken1},
\begin{equation}
 T_0^3\tau_0=T^3\tau=T_f^3\tau_f
 \label{eq16}
\end{equation}

No transverse expansion of the medium considered in the present
calculation which we feel may not be
of much effect when a very high momentum probe
particle is considered and the calculation is done for
charm rapidity, $|y|\simeq 0.0$ 
considering the medium fluid to be always 
on local rest frame~\cite{long1,long2}.

The density of the system is roughly assumed to be,

%The temperature of the plasma at a time $\tau$, assuming
%a chemically equilibrated plasma can be expressed as,
%\begin{equation}
% T(\tau)=\left(\frac{\pi^2}{1.202}\frac{\rho(\tau)}{(9N_f+16)}\right)
%\end{equation}
%where
\begin{equation}
 \rho_g(\tau)=\frac{1}{\pi R^2\tau}\frac{dN}{dy}
 \label{eq17}
\end{equation}

The charged particle multiplicity, '$dN/dy$ is taken from~\cite{multi1}
and multiplied by factors 1.5 get $dN_g/dy$.

After the energy/momentum loss, we can fragment the charm
momentum both from AA and pp
collisions into D-mesons as
D mesons data are readily verifiable from experiments. Schematically,
this can be shown as,
\begin{equation}
E\frac{d^3\sigma}{d^3p}=E_Q\frac{d^3\sigma}{d^3p_Q}\otimes D(Q\rightarrow H_M)~,
\label{eq18}
\end{equation}
where  the fragmentation of the heavy quark $Q$ into the heavy-meson $H_Q$ is
described by the function $D(z)$. We have assumed that the shape of
$D(z)$, where $z=p_D/p_c$, is identical for all the $D$-mesons and
is given by;
\begin{equation}
D^{(c)}_D(z)=\frac{n_D}{z[1-1/z-\epsilon_p/(1-z)]^2}~,
\label{eq19}
\end{equation}
$\epsilon_p$ is the Peterson parameter $\simeq$0.13 and
is taken from~\cite{peterson1}.
The normalization condition satisfied by the fragmentation
function is;
\begin{equation}
\int_0^1 \, dz \, D(z)=1~.
\label{eq20}
\end{equation}

With these energy loss mechanisms in hand, we move
over to
next section calculating charm quark observables.

\begin{figure}
%\begin{center}
\includegraphics[scale=0.30,angle=270]{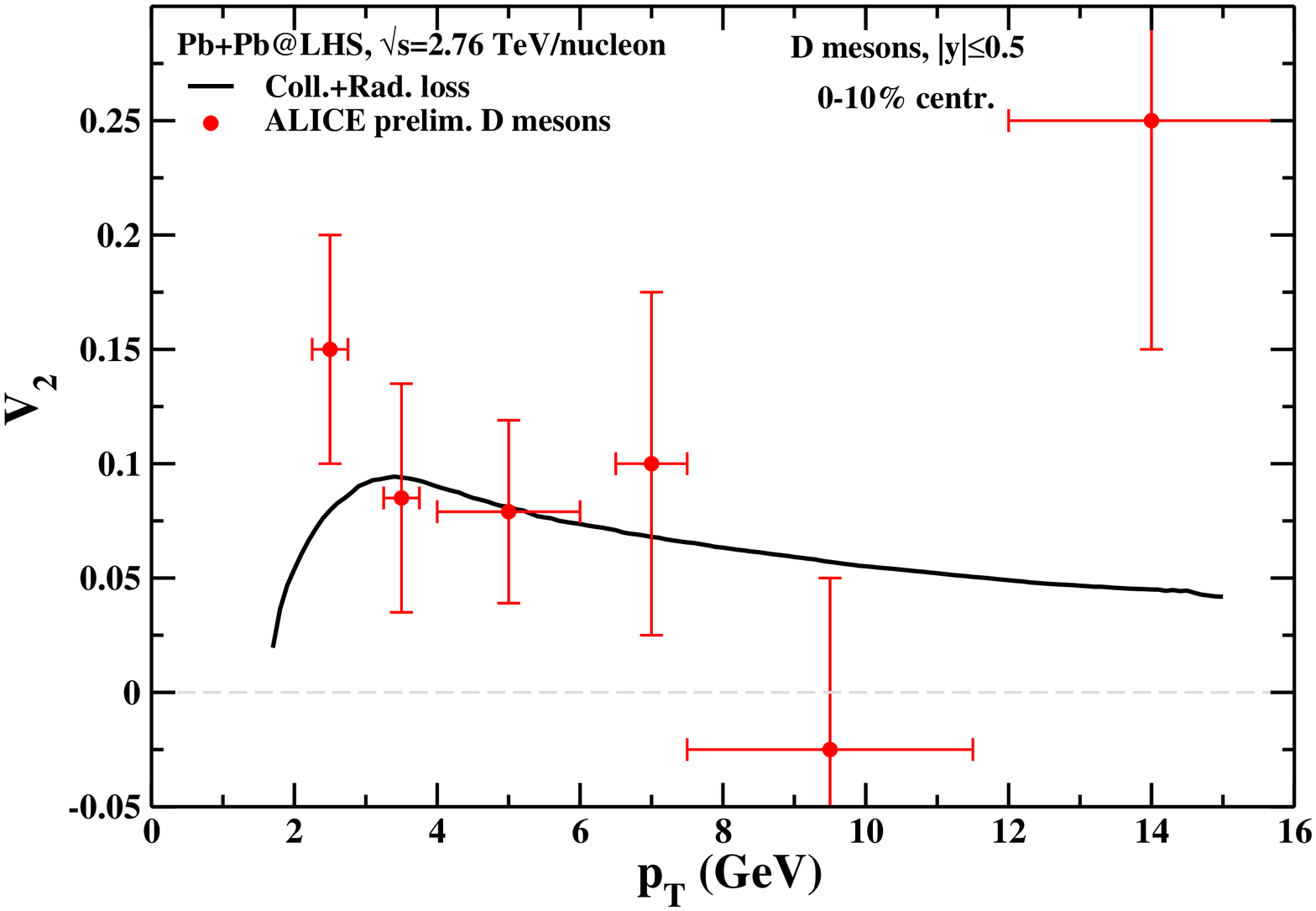}
\includegraphics[scale=0.30,angle=270]{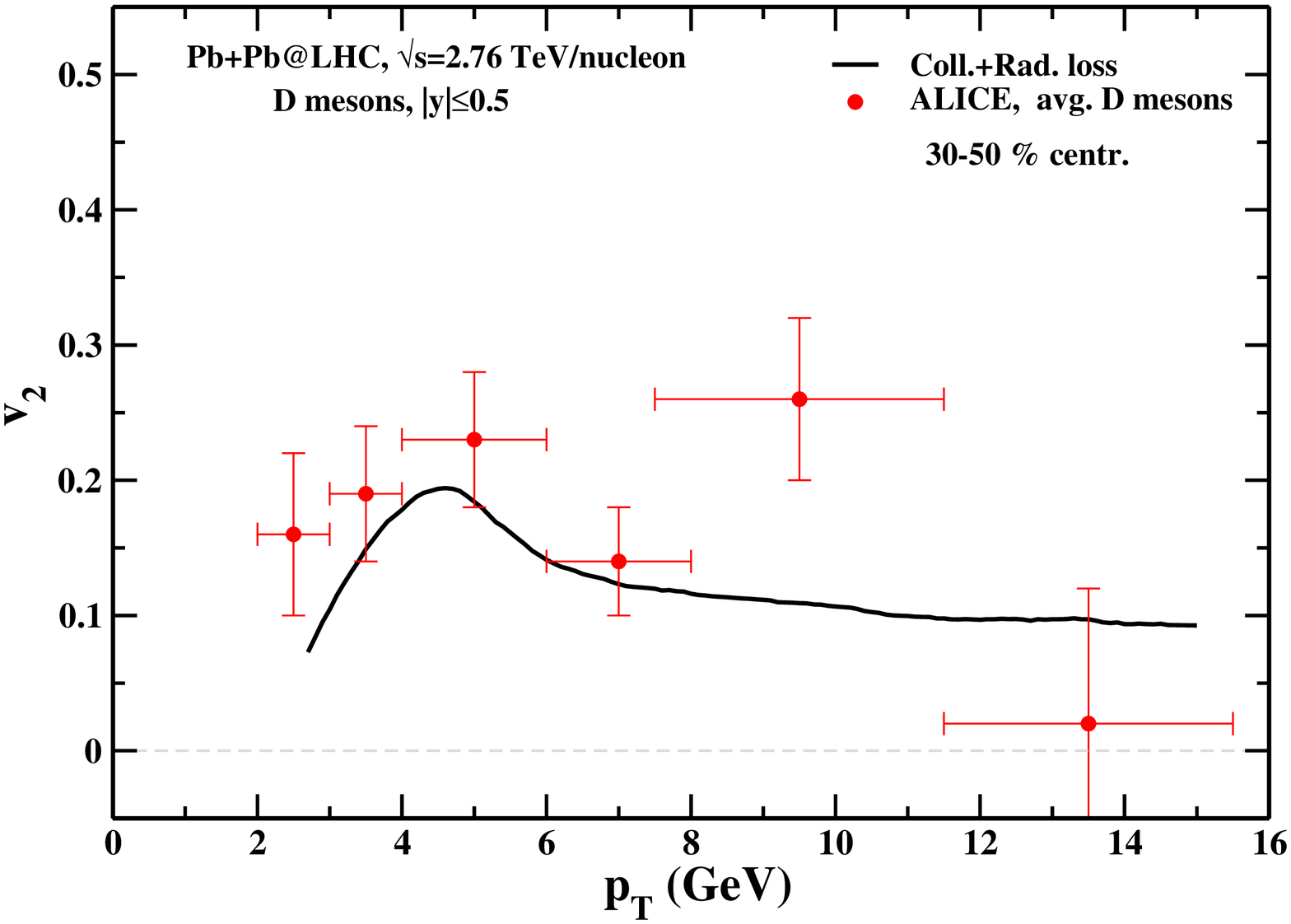}
%\end{center}
\caption{Azimuthal anisotropy, '$v_2$', (left)0-10\% centr. and
(right)30-50\% centr.}
\label{fig2v2}
\end{figure}
section{Observables}

\subsection{\bf {Nuclear Modification Factor}}
The energy loss of quarks(light and heavy) and gluons
is easily observed 
via the suppressed production of hadrons measured
using nuclear modification factor, $R_{AA}$:

\begin{equation}
R_{AA}(p_{T},y)=\frac{dN_{AA}/d^{2}p_{T}dy}
{\langle T_{AA}\rangle d\sigma_{pp}/d^{2}p_{T}dy}
\label{eq21}
\end{equation}
where $N_{AA}$ is the hadron production for the nucleus-nucleus system at
a given impact parameter, $T_{AA}$ is the corresponding nuclear thickness,
and $\sigma_{pp}$ is the cross-section for the production of hadrons at the
coressponding centre of mass energy/mucleon.

\subsection{\bf {Azimuthal Anisotropy}}
Non-central collisions of identical nuclei will lead to
an oval overlap zone, whose length in and out of the reaction plain
would be different. Thus, charm quarks traversing the QGP in and out
of the plain will cover different path lengths and lose different
amount of energy. This would lead to an azimuthal dependence in the
distribution of resulting charmed mesons, whose azimuthal anisotropy
could be measured in terms of the $v_2$ coefficient defined by
\begin{equation}
v_{2}(p_T)=\frac{\int{d\phi \frac{dN}{p_{T}dp_{T}d\phi}\cos{(2\phi)}}}{\int{d\phi \frac{dN}{p_{T}dp_{T}d\phi}}}
\label{eq22}
\end{equation}
where $\phi$ is the azimuthal angle in transverse momentum plane.

\section{Results and discussion}
To begin with the discussion, let me state
that the present calculation assumes a simple phenomenological
model for implementing LPM like effects in heavy quark radiation.
This is different from many earlier calculations such as DGLV, GWP,
BDMPS-Z, or ASW~\cite{rad1,rad2},\cite{rad3}--\cite{rad5} where rigorous calculations and implementations
of Bethe-Heitler(BH)~\cite{BH1} and LPM effects were done. Furthermore there are differences
in the various kinematical cuts between these formalisms and the present
calculation. However in the literature (see Ref.~\cite{hqw3}) the authors have
stated that there will be marginal differences between BH and LPM like effects.
All these differences and comparisons will feature in our future
publications. In this paper I intend to present calculations
on nuclear modification factor, '$R_{AA}$' and azimuthal anisotropy,
'$v_2$' simultaneously for D mesons compared to recent experimental data from LHC.
Similar attempts have been done recently in works by~\cite{raav21,raav22}.

Let us recall that I have taken LO cross-section for the
charm production and multiplied it by a K-factor of 2.5 for NLO
approximation. This factor gets canceled out when taking the
ratios such as $R_{AA}$ and $v_2$. It is known
that dependence on NLO processes
changes the magnitude and shape of the
$p_T$ distribution of charm cross-section
beyond $p_T>$20 GeV when compared to LO spectrum,
and may affect the magnitude of $R_{AA}$ and $v_2$ at high $p_T$ region. We
will be using NLO cross-section for charm production in our future publications.
We have also used coupling '$\alpha_s$'=0.3,
thermalization temperature, $T_0$=450 MeV, and
freeze-out temperature, $T_f$=165 MeV respectively
as the some of the parameters. The transverse expansion of the
medium has been neglected as a first approximation.

In Fig.~\ref{fig1raa}, we plot three figures of nuclear modification factor, $R_{AA}$,
of D mesons, each corresponding to different centrality classes.
The results
are shown for Pb on Pb collision at LHC, $\sqrt{s}$= 2.76 TeV/nucleon.
The data are taken from a recent ALICE prelim. data paper~\cite{dat3} on D mesons.
The effect on $R_{AA}$ due to collisional loss, radiative loss and
total loss are shown separately in each of the plots. We notice
in all the three plots, there is a particular competition
between collisional and radiative effects on $R_{AA}$.
At $p_T<$ 4.0 GeV, collisional loss seems to generate
more suppression while for $p_T >$ 4.0 GeV, radiation
gives more suppression to charm. But neither collisional
loss nor radiative loss can explain the data individually and
in any case a combination of the two is needed for proper
explanation of the observed results. In this paper 
a simple addition of collisional and radiative 
loss is considered although the respective 
mechanisms belong to two separate models. The author feels
that if average energy loss as function of probe charm
energies or path length dependence of the energy loss
is calculated then the difference between collisional
and radiative mechanisms can be shown more clearly.
These features will be addresses in my upcoming publication.

Let us now returns to the results again.
In all three plots we may notice that our calculation of $R_{AA}$
involving both collisional and radiative energy loss can explain
the data within error bars. We may also notice that the heavy quarks
are suppressed more in the mid-$p_T$ region 4.0 GeV $< p_T <$ 10.0 GeV
than rest of the spectrum.
Although there is suppression in the high $p_T$ region,
since mass of the charm quark is negligible
compared to its momentum, it should behave similar to
that of light quarks and gluons and there it shows a rising trend
but more towards saturation much below unity.

Let us now discuss the plots in detail.
The plot (top left) of Fig.~\ref{fig1raa} is our model calculation
for 0-10\% centrality collision (0-7.5\% centr. for data). We find that
the model calculation shows more suppression than the experimental points,
for 2.0 GeV$<p_T<$4.0 GeV and rises steeply for $p_T\leq$ 2.0 GeV where it is
unable to explain the data. But the model is able to explain the experimental
points above 4.0 GeV charm transverse momentum, within the error bars.
Our calculations are based on pQCD techniques and applied here for charm
quark jets. Hence the present calculation may not be able explain
results in the low $p_T$ region where non-perturbative techniques may
have considerable effects.

In plot (Top right) of Fig.~\ref{fig1raa}, which is for 0-20\% centr.
class, the model calculation is able to
explain results very well for entire $p_T$ range shown. However
we feel that for $p_T>$ 14.0 GeV, since the rise is less, it might underestimate
the data. This may be due to the fact that we have considered very large
emitted gluon energy which might have contributed to the more suppression
than expected or evident from experiments. Another reason behind this
apparent mismatch between the present calculation and the data may
be due to the absence of NLO processes or higher order corrections
to the scattering amplitudes which may have some effects on
the suppression.

Also the present calculation assumes eikonal apporximation for
the heavy quark propagator in the medium. If certain non-eikonal
approximation~\cite{tb1} is considered, the effects of soft gluon bremsstrahlung at
various $p_T$ regime might change.
Similarly transverse expansion of the medium might
also bring in some considerable effects which is absent in the
present calculations. All these features are currently under study
and will be reported in my future publications.

Also in plot (bottom) of Fig.~\ref{fig1raa}, we find the rise is even lesser
and grossly underestimate the data around 12.0 GeV$<p_T<$14.0 GeV.
The reason here as discussed last may be again due
to large emitted gluon energy. However in all the three plots, the model
calculation explains the data to certain extent within
error bars. While for low transverse momentum region,
it fails as other non-perturbative
effects etc. must be incorporated.

Now let us move onto Fig.~\ref{fig2v2} where we have shown two
plots of azimuthal anisotropy $v_2$ of D mesons, for
two different centrality classes. Let us recall
that our calculation only consists of energy
loss by charm quarks and no transverse expansion
of the medium has been included. In plot (left)
of Fig.~\ref{fig2v2} we have $v_2$ of D mesons for 0-10\% centr.
Since this is similar to the most central collision
although not identical to head on collisions, we expect a very
small elliptic flow to form and also a very small
variation with charm momentum (almost flat). This
is also evident from our model calculation. In plot (right)
of the same Fig.~\ref{fig2v2}, we have 30-50\% centrality where
ellipticity is more pronounced and we have more prominent
$v_2$. Our model calculation roughly explains the data
but also underestimates it. We feel that absence of any
transverse expansion scenario in our model is the reason
behind the difference between calculated results and the 
observed data.

\section{Summary and conclusion}
In this paper we have made an attempt in explaining
D mesons data at LHC, $\sqrt{s}=$2.76 TeV/nucleon.
We have calculated charm $p_T$ distribution at
leading order multiplied by a k factor. We have included
heavy quark collisional and radiative energy loss and attempted
to calculate collisional and radiatve charm energy loss per unit
path length. The LPM effect for the radiative processes is
included heuristically. Consequently $R_{AA}$ and $v_2$ are
calculated for different centrality classes and compared
with some of the recent D mesons experimental results from ALICE.
In future we will attempt to incorporate Non-eikonal approximation, soft gluon
bremsstrahlung as well as for $\mu\neq 0$, baryonic chemical potential.
We would like to include other thermal models or hydrodynamical
calculations for an expanding medium along transverse direction 
to give this calculation a more realistic picture.
Further extension of this work will also include the calculations of
other transport coefficients such as momentum broadening, $\hat{q}$,
drag  and diffusion coefficients of charm and their momentum,
temperature, and path length dependencies.

\section*{Acknowledgments}
One of us (MY) would like to thank Dr. sanjay K. Ghosh, Dr. Rajarshi Ray, Bose Institute, and
Dr. Trambak Bhattacharyya, IIT, Indore, and Dr. Surasree Mazumder,
IOP Bhubaneshwar, for the discussions related
to the present work.


\begin{thebibliography}{99}
\bibitem{qgp1}J.~C.~Collins, and M.~J.~Perry, Phys. \ Rev. \ Lett.
              {\bf 34}, 1353 (1975);
              R.~Hagedorn, and J.~Rafelski, Phys. \ Lett.
              {\bf B 97}, 136 (1980);
              L.~D.~McLerran, and B.~Svetitsky, Phys. \ Lett.
              {\bf B 98}, 195 (1981).
\bibitem{qgp2}K.~Kajantie, C.~Montonen and C.~Pietarinen,
              Zeit. \ Phys. \ {\bf C 9}, 253 (1981);
              R.~Hagedorn and J.~Rafelski,
              Phys. \ Lett. \ {\bf B 97}, 180 (1980).
\bibitem{qgp3}J.~W.~Harris and B.~M\"{u}ller,
              Ann. \ Rev. \ Nucl. \ Part. \ Sci. {\bf 41}, 96 (1996);
              B.~M\"{u}ller, Rep. \ Prog. \ Phys. {\bf 58}, 611 (1998).
              S.~A.~Bass, M.~Gyulassy, H.~St\"{o}cker and W.~Greiner,
              J. \ Phys. \ {\bf G}: Nucl. \ Part. \ Phys. {\bf 25}, R1 (1999).                            
\bibitem{sig1}M.~Gyulassy, I.~Vitev, and X.~N.~Wang, Phys. \ Rev. \ Lett.
              {\bf 86}, 2537 (2001);
\bibitem{sig2}S.~Wicks, W.~Horowitz, M.~Djordjevic and M.~Gyulassy, Nucl.\ Phys.\ A
              {\bf 783}, 493 (2007)
\bibitem{sig3}X.-N.~Wang, Phys. \ Rev. {\bf C 63}, 054902 (2001);
              J.~Adams et al. (STAR collaboration), Phys. \ Rev. \ Lett.
              {\bf 91}, 172302 (2003).
\bibitem{sig4}R.~J.~Fries, S.~A.~Bass, and B.~Muller, Phys. \ Rev. \ Lett.
              {\bf 94}, 122301 (2005).
\bibitem{sig5}T.~Matsui, and H.~Satz, Phys. \ Lett.
              {\bf 178}, 416 (1986).
\bibitem{pro1}Z.-W.~Lin, and M.~Gyulassy, Phys. \ Rev.
              {\bf C 51}, 2177 (1995); {\bf C 52}, 440 (1995) (erratum);
              M.~Younus, and D.~K.~Srivastava, J. \ Phys. \ {\bf G}: \ Nucl. \ Part. \ Phys.
              {\bf 37}, 115006 (2010).
\bibitem{pro2}R.~Vogt,
              J.\ Phys.\ Conf.\ Ser.\  {\bf 509}, 012007 (2014);
              R.~Vogt,
              Indian J.\ Phys.\  {\bf 85}, 1075 (2011).
\bibitem{suppress1}A.~Drees, Nucl. \ Phys. \ {\bf A 698}, 331 (2002);
                   E.~Shuryak, Nucl. \ Phys. \ {\bf A 750}, 64 (2005);
                   S.~Jeon and G.~D.~Moore, Phys. \ Rev. \ {\bf C 71}, 034901 (2005).
\bibitem{suppress2}X-N.~Wang, Nucl. \ Phys. \ {\bf A 750}, 98 (2005);
                   A.~K.~Chaudhuri, Phys. \ Lett. \ {\bf B 659}, 531 (2008);
                   David d'Enterria and B.~Betz, Lect. \ Notes \ Phys. \ {\bf 785}, 285 (2010).              
\bibitem{dat1}S.~S.~Adler {\it et al.} (PHENIX Collaboration), Phys. \ Rev. \ Lett.
              {\bf 98}, 172301 (2007);
              B.~I.~Abelev {\it et al.} (STAR Collaboration), Phys. \ Rev. \ Lett.
              {\bf 98}, 192301 (2007).
\bibitem{dat2}K.~Aamodt et al. (ALICE Collaboration), Phys. \ Rev. \ Lett.
              {\bf 105}, 252301 (2010); Phys. \ Rev. \ Lett. {\bf 105}, 252302 (2010).
\bibitem{dat3}B.~B.~Abelev {\it et al.}  [ALICE Collaboration],
              Phys.\ Rev.\ C {\bf 90}, no. 3, 034904 (2014);
              A.~Festanti [ALICE Collaboration],
              Nuovo Cim.\ C {\bf 037}, no. 01, 287 (2014).
\bibitem{therm1}G.~D.~Moore and D.~Teaney,
                Phys.\ Rev.\ C {\bf 71}, 064904 (2005);
                K.~Dusling, G.~D.~Moore and D.~Teaney,
                Phys.\ Rev.\ C {\bf 81}, 034907 (2010).
\bibitem{hqw1}S.~Wicks {\it et al.}, Nucl. \ Phys. {\bf A 784}, 426 (2007); ibid. {\bf 872}, 265 (2011);
              H.~van~Hees {\it et al.}, Phys. \ Rev. {\bf C 73}, 034913 (2006);
              M.~He {\it et al.}, Nucl. \ Phys. {\bf A 910}, 409 (2013).
\bibitem{hqw2}S.~Cao {\it et al.}, Nucl. \ Phys. {\bf A 904}, 653 (2013);
              Phys. \ Rev. {\bf C 88}, 044907 (2013);
              M.~Nahrgang {\it et al.}, Nucl. \ Phys. {\bf A 904}, 992c (2013);
              Phys. \ Rev. {\bf C 91}, 014904 (2015);
              Nucl. \ Phys. {\bf A 910}, 301 (2013);
\bibitem{hqw2b}J.~Uphoff {\it et al.}, Nucl.\ Phys. {\bf A 931}, 535 (2014),
              Nucl. \ Phys. {\bf A 931}, 937 (2014).
\bibitem{hqw3}S.~Mazumder {\it et al.}, Phys. \ Rev. {\bf C 84}, 044901 (2011);
              S.~Das {\it et al.}, Phys. \ Rev. {\bf C 90}, 044901 (2014);
              R.~Abir {\it et al.}, Phys. \ Lett. {\bf B 715}, 183 (2012).
\bibitem{hqw4}M.~Younus {\it et al.},
              J. \ Phys. {\bf G}: Nucl. Part. Phys. {\bf 39}, 095003 (2012);
              M.~Younus, C.~E.~Coleman-Smith, S.~A.~Bass and D.~K.~Srivastava,
              Phys. \ Rev. \ {\bf C 91}, 024912 (2015).
\bibitem{hqc1}X.~Zhu, N.~Xu, and P.~Zhuang, 
              J. \ Phys. \ {\bf G}: \ Nucl. \ Part. \ Phys. {\bf 36}, 064025 (2009); 
              M.~Younus {\bf et al.}, 
              J. \ Phys. \ {\bf G}: \ Nucl. \ Part. \ Phys. {\bf 39}, 025001 (2012);
              M.~Younus {\bf et al.}, 
              J. \ Phys. \ {\bf G}: \ Nucl. \ Part. \ Phys. {\bf 40}, 065004 (2013).
\bibitem{hqw5}F.~D'Eramo, H.~Liu and K.~Rajagopal, Phys. \ Rev. \ {\bf D 84}, 06015 (2011);
              A.~Majumder, B.~M\"{u}ller and S.~Mr\'{o}wczy\'{n}ski, Phy. \ Rev. \ {\bf D 80}, 125020 (2009);
              R.~Baier, D.~Schiff and B.~G.~Zakharov, Annu. \ Rev. \ Nucl. \ Part. \ Sci. {\bf 50}, 37 (2000);
              A. ~Kovner and U.~A.~Wiedemann, arXiv:hep-ph/0304152v1 2003;
              P.~Arnold and W.~Xiao, Phys. \ Rev. \ {\bf D 78}, 125008 (2008).
\bibitem{pro2}E.~Eichten, I.~Hinchliffe, K.~D.~Lane, and C.~Quigg,
              Rev. \ Mod. \ Phys. {\bf 56}, 579 (1984)
              E.~Eichten {\it et al.}, Rev. \ Mod. \ Phys. {\bf 58}, 1065 (1986)(addendum);
              U.~Jamil, and D.~K.~Srivastava, J. \ Phys. \ {\bf G}: Nucl. \ Part. \ Phys. {\bf 37} 085106 (2010).
\bibitem{pdf1}H.~L.~Lai {\it et al.}, Phys. \ Rev. \ {\bf D 55}, 1280 (1997).
\bibitem{sdf1}K.~J.~Eskola, V.~J.~Kolhinen, and C.~A.~Salgado,
              Euro. \ J. \ Phys. {\bf C 9}, 61 (1999);
              K.~J.~Eskola, V.~J.~Kolhinen, and P.~V.~Ruuskanen,
              Nucl. \ Phys. \ {\bf B 535}, 351 (1998).
\bibitem{coll1}M.~Thoma and M.~Gyulassy,
               Nucl. \ Phys. {\bf B 351}, 491 (1991);
               P.~B.~Gossiaux and J.~Aichelin,
               Phys. \ Rev. {\bf C 78}, 014904 (2008).
\bibitem{coll2}A.~Peshier and S.~Peigne,
               Phys. \ Rev. {\bf D 77}, 114017 (2008).
\bibitem{pro3}B.~L.~Combridge, Nucl. \ Phys. \ {\bf B 151}, 429 (1979).
\bibitem{rad1}R.~Baier {\it et al.},
              Nucl. \ Phys. {\bf B 478}, 577 (1996),
              Nucl. \ Phys. \ {\bf B 483}, 291 (1997);
              M.~Gyulassy. P.~Levai and I.~Vitev,
              Phys. \ Rev. \ Lett. {\bf 85}, no. 26, 5535 (2000).
\bibitem{rad2}M.~G.~Mustafa, D.~Pal and D.~K.~Srivastava,
              Phys. \ Rev. {\bf C 57}, 889 (1998).
\bibitem{form1}R.~Baier {\it et al.}, 
               Phys. \ Lett. {\bf B 345}, 277 (1995).
\bibitem{LPM1}L.~D.~Landau and I.~Pomeranchuk,
              Dokl. \ Akad. \ Nauk. \ Ser. \ Fiz. {\bf 92}, 535 (1953);
              {\bf 92}, 735 (1953) (In Russian);
              A.~B.~Migdal, Phys. \ Rev. {\bf 103}, 1811 (1956).
\bibitem{DK1}Y.~L.~Dokhshitzer, and D.~Kharzeev,
             Phys. \ lett. \ {\bf b 519}, 199 (2001).
\bibitem{gb1}J.~F.~Gunion and G.~Bertsch,
             Phys. \ Rev. \ {\bf D 25}, no. 3, 746 (1982).
\bibitem{abir1}R.~Abir, C.~Greiner, M.~Martinez, and M.~G.~Mustafa,
               Phys. \ Rev. \ {\bf D 83}, 011501(R) (2011).
\bibitem{bjorken1}J.~D.~Bjorken,
                  Phys.\ Rev.\ D {\bf 27}, 140 (1983).
\bibitem{long1}R.~Vogt, B.~V.~Jacak, P.~L.~MvGaughey, and P.~V.~Russkanen
               Phys. \ Rev. {\bf D 49}, 3345 (1994).
\bibitem{long2}S.~Gavin, P.~L.~Mcgaughey, P.~v.~Ruuskanen, and R.~Vogt,
               Phys. \ Rev. {\bf C 54},2606 (1996);
               S.~Sarkar {\it et al.}, 
               Phys. \ Rev. {\bf C 51}, 318 (1995); 2845 (1995) (Erratum).
\bibitem{multi1}D.~Kharzeev, E.~Levin, and M.~Nardi,
                Nucl. \ Phys. {\bf A 747}, 609 (2005).
\bibitem{peterson1}C.~Peterson, D.~Schlatter, I.~Schmitt, and P.~Zerwas,
                   Phys. \ Rev. \ {\bf D 27}, 105 (1983).
\bibitem{rad3}N.~Armesto, C.~A.~Salgado, and U.~A.~Wiedemann,
              Phys. \ Rev. {\bf D 69}, 114003 (2004).
\bibitem{rad4}M.~Gyulassy and X.-N.~Wang,
              Nucl. \ Phys. \ {\bf B 420}, 583 (1994).
\bibitem{rad5}X.-N.~Wang, M.~Gyulassy, and M.~Plummer,
              Phys. \ Rev. {\bf D 51}, no. 7, 3436 (1995).
\bibitem{BH1}H.~Bethe and W.~Heitler, Proc. \ Royal Soc. \ London,
             series A, {\bf 46}, 83 (1934);
             L.~I.~Schiff, Phys. \ Rev. \ {\bf 83}, 252 (1951);
             R.~Baier, Y.~L.~Dokhshitzer, S.~Peigne and L.~I.~Schiff,
             Phys. \ Lett. \ {\bf B 345}, 277 (1995).              
\bibitem{raav21}S.~Das {\it et al.},
                arXiv:1502.03757 [nucl-th] 2015.
\bibitem{raav22}M.~Nahrgang {\it et al.},
                Nucl. \ Phys. \ {\bf A 931}, 575 (2014).
\bibitem{tb1}T.~Bhattacharyya {\it et al.}
             arXiv:1307.6931 [hep-ph] 2013.

\end{thebibliography}
\end{document}